\documentclass[prd,preprint,superscriptaddress]{revtex4}
\usepackage{graphicx}
\usepackage{epsfig}
\usepackage{amssymb}
\newcommand{\ochi}{\Omega_\chi}
\newcommand{\odm}{\Omega_{DM}}

\newcommand{\gev}{~\mathrm{GeV}}
\newcommand{\tev}{~\mathrm{TeV}}
\def\lsim{\mathrel{\hbox{\rlap{\hbox{\lower4pt\hbox{$\sim$}}}\hbox{$<$}}}}

\begin{document}

\title{Direct detection of neutralino dark matter  in non-standard
  cosmologies}

\author{Graciela B. Gelmini}
\email{gelmini@physics.ucla.edu}
\affiliation{Department of Physics and Astronomy, UCLA, 475 Portola Plaza, Los
  Angeles, CA 90095, USA}

\author{Paolo Gondolo}
\email{paolo@physics.utah.edu}
\affiliation{Department of Physics, University of Utah, 115 S 1400 E \# 201,
  Salt Lake City, UT 84112, USA}

\author{Adrian Soldatenko}
\email{asold@physics.ucla.edu}
\affiliation{Department of Physics and Astronomy, UCLA, 475 Portola Plaza, Los
  Angeles, CA 90095, USA}

\author{Carlos E. Yaguna}
 \email{yaguna@physics.ucla.edu}
\affiliation{Department of Physics and Astronomy, UCLA, 475 Portola Plaza, Los
  Angeles, CA 90095, USA}


\begin{abstract}
We compute the neutralino direct detection rate in non-standard cosmological
scenarios where neutralinos account for the dark matter of the
Universe. Significant differences are found when such rates are compared with
those predicted by the standard cosmological model. For bino-like neutralinos,
the main feature is the presence of additional light ($m_\chi\lesssim 40\gev$) and heavy
($m_\chi\gtrsim 600\gev$) neutralinos with detection rates  within the sensitivity of
future dark matter experiments. For higgsino- and wino-like neutralinos lighter than $m_\chi \sim 1\tev$, enhancements of more
than two orders of magnitude in the largest detection rates are
observed. Thus, if dark matter is made up of neutralinos, the prospects for
their direct detection are in general more promising than in the standard cosmology.
\end{abstract}


\maketitle
\section{Introduction}
The Large Hadron Collider is now in its final preparation stages and may soon be searching
for supersymmetric particles. Among them, the lightest neutralino in the minimal supersymmetric standard model
plays a distinctive role as a dark matter candidate \cite{Griest:2000kj}. It is neutral, weakly
interacting, and stable (provided it is the lightest supersymmetric
particle). If evidence for low energy supersymmetry is found, it will strongly support the idea
that neutralinos constitute the dark matter of the Universe. A logical next
step would then be the use of neutralinos as cosmological probes of the early
Universe. Neutralinos could, in particular, test the standard cosmological model well before big bang
nucleosynthesis. Being an observable sensitive to
the conditions in the early Universe, the neutralino
direct detection rate provides a plausible way of discriminating between different
cosmological models, and therefore an indirect way of testing the standard scenario. Most studies on the direct detection of
neutralinos already assume the standard cosmology so it is not known what to
expect in a more general cosmological framework.

The vastness of the supersymmetric parameter space is the most compelling
reason to assume the standard cosmological model. In a general setup,
neither the neutralino mass or gauge composition nor its interaction rate, for
example,  can be determined a priori. To reduce such uncertainties, the dark matter constraint is
usually imposed on supersymmetric models. That is, the neutralino relic density is computed within the
standard cosmological model and only models with $\Omega_{\rm std}<\Omega_{\rm DM}$ are
considered (here $\Omega_{\rm std}$ is the neutralino density in the standard cosmological model, and $\Omega_{\rm DM}$ is the cold dark matter density, both in units of the critical density). This bound, it turns out, is very effective in restricting the parameter
space of supersymmetric models. In minimal supergravity models (mSUGRA), for
instance, the neutralino typically has a 
small annihilation rate in the early Universe, thus its relic density  tends to be
larger than observed. At the end, the requirement $\Omega_{\rm std} < \Omega_{\rm DM}$ is found to  be satisfied only along  four narrow regions: the ``bulk" (with a light neutralino and tight
accelerator constraints), the ``coannihilation region" (where the stau is
almost degenerate with the neutralino and coannihilation effects suppress the
relic density), the ``funnel region" (where $m_\chi\simeq m_A/2$ and resonance
effects enhance the $\chi$-$\chi$ annihilation rate) and the ``focus point
region" (where the neutralino acquires a non-negligible higgsino fraction). Accounting for the dark
matter provides, in fact, the  most stringent constraint on supersymmetric models, well over  precision data or accelerator searches (see e.g.~\cite{regions}).

Though useful in reducing the supersymmetric parameter space, the dark matter
constraint should not be taken for granted, as it  relies on 
untested assumptions about the early Universe. In 
particular, it postulates that the entropy of matter and radiation is  conserved
and that the Universe is radiation dominated at high
temperatures ($T\sim m_\chi$). Several scenarios where such assumptions
do not hold and, more generally, where the evolution of the  
Universe before big bang nucleosynthesis deviates from the standard
cosmological model, have been studied in the literature. They are generically
known as 
non-standard cosmologies and  include models with  gravitino~\cite{gravitino},
moduli~\cite{Moroi-Randall} or Q-ball decay \cite{Fujii}, 
thermal inflation \cite{Lazarides}, the Brans-Dicke-Jordan~\cite{Brans-Dicke-Jordan}
cosmological model, models with anisotropic expansion~\cite{anisotropic} or
quintessence domination~\cite{quintessence}. Non-standard cosmological models
are viable alternatives against which the predictions of the standard scenario
may be compared.

In non-standard cosmological scenarios, the neutralino relic density
$\Omega_\chi$  may be larger or smaller than $\Omega_{\rm std}$
\cite{kamionkowski-turner}-\cite{ggsy}. Smaller
densities are usually the result of an episode of entropy production that
dilutes the neutralino abundance. Larger densities  are
due either to additional contributions to the expansion
rate of the Universe, or to non-thermal neutralino production
mechanisms. Usually these scenarios contain additional parameters that can be adjusted to modify
the neutralino relic density. \textbf{A distinctive feature of non-standard
cosmologies is that the new physics they incorporate does not manifest in accelerator or detection
experiments. That is certainly the case, for instance, for the several models mentioned
above. Neutralino scattering rates, therefore, are not affected by the
cosmological model.}

\textbf{A prototype non-standard cosmological model 
is that of a scalar field $\phi$ with couplings of gravitational strength whose late decay reheats the Universe to a low reheating
temperature. The reheating temperature in this scenario can be lower than the standard  neutralino freeze-out temperature without
spoiling primordial nucleosynthesis~\cite{hannestad}.  Such scalar fields are  common 
in superstring models where they appear as moduli fields.  In these models, the decay of
$\phi$ into radiation increases the entropy, diluting the
neutralino number density.  Instead, the decay of $\phi$ into supersymmetric
particles, which eventually decay into neutralinos, increases the neutralino
number density. In this non-standard cosmological model it has been shown that practically all 
 neutralinos   can have the density
of the dark matter, provided the 
  right combination of   two parameters can be achieved in the high energy theory:
 the reheating temperature, and the ratio of the
 number of neutralinos produced per $\phi$ decay over the $\phi$ field mass~\cite{gg, ggsy}.}

In this paper, we compute the neutralino direct detection rate in generic
cosmological scenarios where neutralinos constitute the dark matter of the
Universe.  That is, we assume that, independently of the supersymmetric spectrum, the parameters of the  non-standard
cosmological model can always be  chosen so that  $\ochi=\odm$. By randomly
scanning the supersymmetric parameter space, we obtain a large sample of
models and compute their detection rates in non-standard cosmologies. These predictions are
then compared with those obtained within the standard cosmological model. Our
goal is twofold. First, we explore the possibility of using the
neutralino direct detection rate as a test of the standard cosmological model.
Second, we establish the potential of future dark matter
detectors in probing the parameter space of supersymmetric models in a
cosmology-independent setup.

 \section{The supersymmetric models}
 In the MSSM, neutralinos are linear combinations of the fermionic partners of the neutral electroweak bosons, called bino ($\tilde B^0$) and wino ($\tilde W_3^0$), and of the fermionic partners of the neutral Higgs bosons, called higgsinos ($\tilde H_u^0$, $\tilde H_d^0$). We assume that the lightest neutralino, $\chi$, is the dark matter candidate. Its composition can be parameterized as
\begin{equation}
\chi=N_{11}\tilde B^0+N_{12}\tilde W_3^0+N_{13}\tilde H_d^0+N_{14}\tilde H_u^0\,.
\label{eq:comp}  
\end{equation}
Because the neutralino interactions are determined by its gauge content, it is
useful to distinguish between bino-like ($N_{11}^2 > N_{12}^2,\,
N_{13}^2+N_{14}^2$), wino-like ($N_{12}^2 > N_{11}^2,\,
N_{13}^2+N_{14}^2$), and higgsino-like ($N_{13}^2 + N_{14}^2>
N_{11}^2,\,N_{12}^2$) neutralinos according to the hierarchy of terms in
(\ref{eq:comp}). This classification implies that even so-called \emph{mixed}
neutralinos, those with two or more comparable components,  are considered as either binos, winos or higgsinos.

Bino-like neutralinos annihilate mainly into fermion-antifermion pairs through
sfermion exchange. Such  annihilation cross-section is helicity suppressed and
gives rise to  a  standard relic density that is usually larger than observed. Agreement with  the observed
dark matter abundance can still be achieved  in standard cosmological
scenarios but only in restricted regions of the parameter space where special
mechanisms such as coannihilations or resonant annihilations help reduce the
relic density. Owing to the gaugino unification condition, bino-like neutralinos are a generic prediction of minimal supergravity models.

Wino-like and higgsino-like neutralinos annihilate mostly into gauge
bosons ($W^+W^-$, $ZZ$, if kinematically allowed) through neutralino or
chargino exchange; otherwise they annihilate into fermions. Due to
coannihilations with the lightest chargino (and, for higgsinos, with  the
next-to-lightest neutralino), their  standard relic density is rather
small. Neutralino masses as large as $1 ~\mathrm{TeV}$ for higgsinos or
$2~\mathrm{TeV}$ for winos are required to bring their thermal density within
the observed range. Wino-like and
higgsino-like neutralinos can be obtained in models with non-universal gaugino
masses; models with anomaly mediated supersymmetry breaking (AMSB) \cite{amsb}, for instance, feature a wino-like neutralino.

We consider a general class of MSSM models  defined in terms of the parameter set
$M_{3}$, $M_2$, $M_1$, $m_A$, $\mu$, $\tan\beta$, $m_{\tilde q}$, $m_{\tilde \ell}$ $A_t$, and
$A_b$. Here $M_{i}$ are the three gaugino masses, $m_A$ is the mass of the
pseudoscalar higgs boson, and $\tan\beta$ denotes the ratio $v_2/v_1$. The
soft breaking scalar masses are defined through the simplifying ansatz
$M_Q=M_U=M_D=m_{\tilde q}$ and $M_E=M_L=m_{\tilde \ell}$, whereas the trilinear couplings are given by
$A_U=\mathrm{diag(0,0,A_t)}$, $A_D=\mathrm{diag(0,0,A_b)}$, and $A_E=0$. All
these parameters are defined at the weak scale.   Specific realizations of
supersymmetry breaking such as mSUGRA, mAMSB \cite{amsb} or split-SUSY \cite{splitsusy} are similar to - though not
necessarily coincide with - particular examples of these models.

We performed a random scan of such parameter space within the following ranges
\begin{eqnarray}
10 ~\mathrm{GeV}&< M_1,M_2,M_3 &< 50 ~\mathrm{TeV}\\
40 ~\mathrm{GeV}&< m_A, \mu, m_{\tilde q}, m_{\tilde \ell} &<50 ~\mathrm{TeV}\\
-3 m_0 &< A_t, A_b&< 3 m_0\\
1&< \tan\beta&<60
\end{eqnarray} 
A logarithmic distribution was used for $M_i$, $m_A$, $\mu$,  $m_{\tilde q}$
and $m_{\tilde \ell}$, and a linear one for $A_t$, $A_b$, and $\tan\beta$; the
sign of $\mu$ was randomly chosen. After imposing accelerator constraints, as contained in
DarkSUSY version 4.1~\cite{Gondolo:2004sc}, a sample of about 
$10^5$ viable models was  obtained. The following  analysis is based on such a sample of
supersymmetric models.

\section{Results}

\begin{figure}
\begin{center}
\includegraphics[scale=0.5]{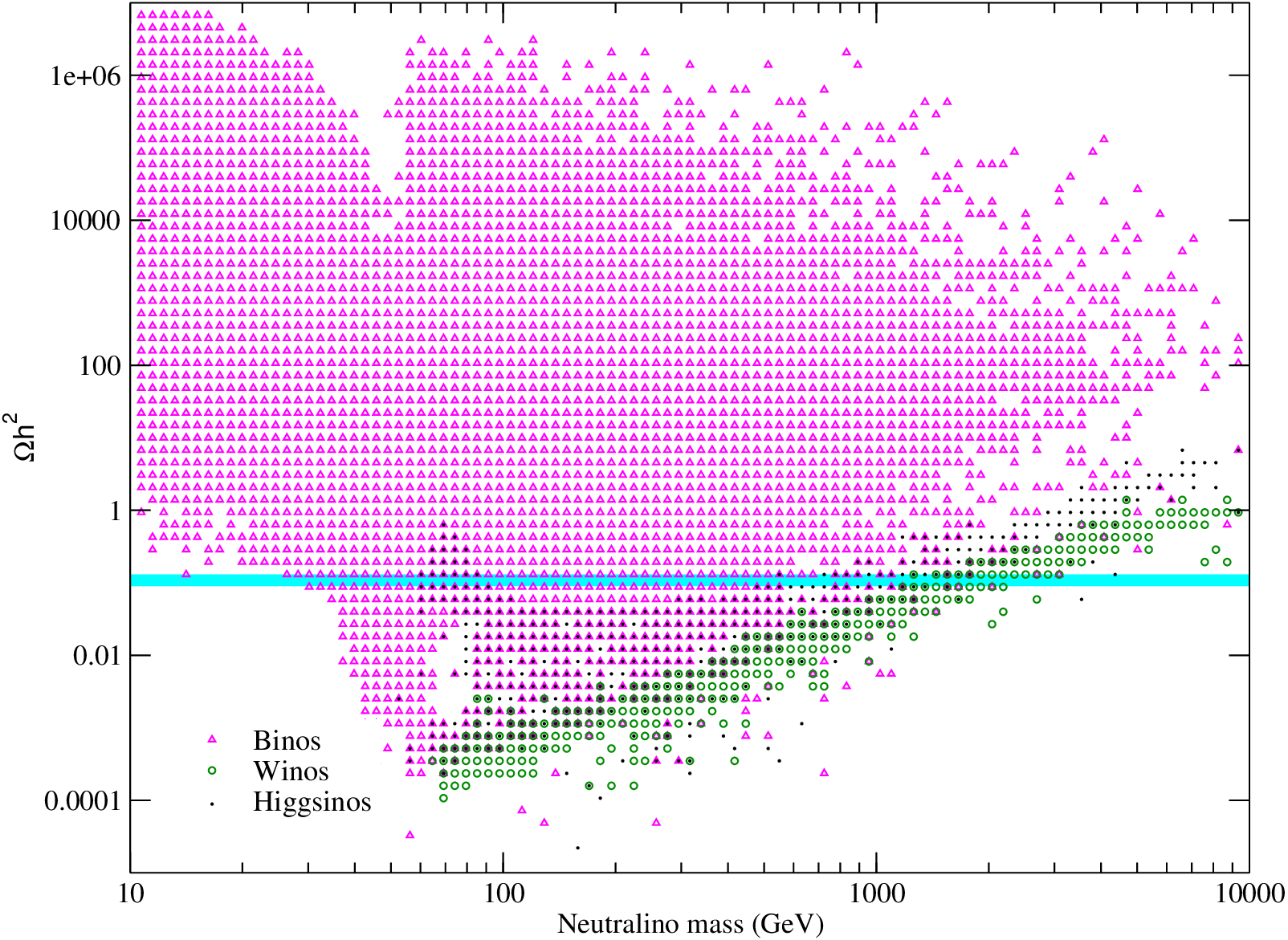}
\caption{The \emph{standard} neutralino relic density as a function of the neutralino mass for
our sample of models. The models are differentiated according to the bino,
wino, or higgsino character of the lightest neutralino. The horizontal band
indicates the dark matter range.}
\label{omega-bwh-MSSM}
\end{center}
\end{figure}
Figure  \ref{omega-bwh-MSSM} shows the \emph{standard} relic density  as a
 function of the neutralino mass for our sample of models. Each cell
 -triangle, circle or dot- represents a small region around which at
 least one model was found. The models are classified as binos, winos, or
 higgsinos, according to the gauge composition of the lightest neutralino. The
 horizontal band corresponds to the observed dark matter density $\Omega_{\rm std}h^2=\Omega_{\rm dm} h^2=$ 
$0.109^{+0.003}_{-0.006}$, obtained for a $\Lambda$CDM model with
 scale-invariant primordial perturbation spectrum through a global fit of
 cosmic microwave background, supernovae, and large scale structure
 data~\cite{wmap2}. Several observations can be made from this figure. Models with bino-like neutralinos are spread over a
 wide area and usually give a rather large relic density. Models with wino-
 and higgsino-like neutralinos, on the contrary, are concentrated over narrow
 bands and their relic density exceeds the dark matter density only for large
 masses, $m_\chi\gtrsim 1\tev$. Finally, notice that in our
 sample the neutralino relic density varies between $10^6$ and
 $10^{-4}$.

We now want to compute, for our set of models, the neutralino interaction
rates in generic cosmologies where the neutralino accounts for the dark matter
and compare them with those obtained in the standard cosmology. Since
spin-dependent searches are harder than spin-independent ones, we will
focus on the latter. The neutralino interaction rate in direct dark matter
detection experiments is  proportional to the 
product of the spin-independent neutralino-nucleus cross section $\sigma_{\rm SI}$ and the number density  of neutralinos 
passing through the detector, $f$. We assume that, as expected  for collisionless cold dark
 matter, $f= \Omega_\chi/ \Omega_{\rm dm}$. $\sigma_{\rm SI}$ is
 determined only by the supersymmetric spectrum but $\Omega_\chi$ is sensitive
 to the cosmological setup. Thus, the neutralino detection rate depends on the
 cosmology only through $f$.

 If the standard cosmological model is assumed, then all models above
the horizontal band in figure \ref{omega-bwh-MSSM} are rejected. They have a
standard relic density larger than the
observed dark matter density ($\Omega_{\rm std}>\Omega_{\rm DM}$) and therefore are
considered incompatible with  cosmological observations. Models with a
relic density below the dark matter density are still viable, though
neutralinos make up only a fraction of the dark matter. They have $f<1$, so
their detection rate is typically suppressed.   Finally, those models with a
neutralino relic density within the observed dark matter range are viable and
have $f=1$. They have been the  focus of the large majority of studies on
neutralino direct detection.

In non-standard cosmologies, $\ochi=\odm$ may be ensured and
the previous picture is modified in two important ways. On the one hand, the
viable parameter space is different. In fact, overdense models, those with
$\Omega_{\rm std}>\Omega_{\rm DM}$, can no longer be rejected. On the other hand,
underdense models, those with $\Omega_{\rm std}<\Omega_{\rm DM}$, no longer will have
the $f<1$ suppression factor in the detection rate. Hence, in non-standard
cosmologies, we expect more viable models and larger detection rates. A
priori, however, it is not possible to predict the detection rate for the
new viable models or to know whether the enhanced detection rates are within
the sensitivity of future dark matter detection experiments. Thus, a
careful analysis is required to establish the implications of non-standard
cosmologies for dark matter searches. In the following, such an analysis will
be carried out.

\begin{figure}
\begin{center}
\includegraphics[scale=0.5]{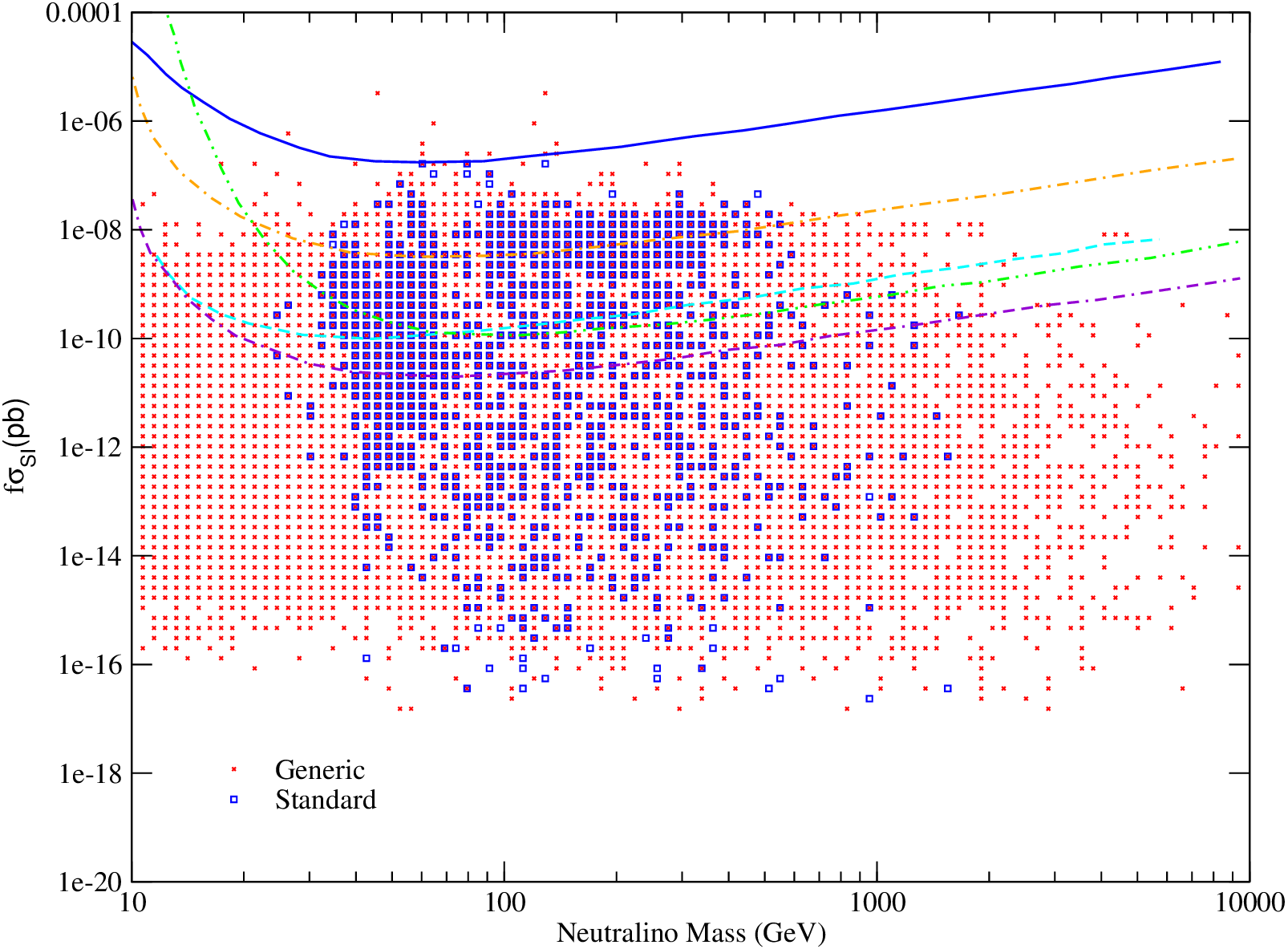}
\caption{Spin-independent neutralino-proton cross section $\sigma_{SI}$ multiplied by the neutralino halo fraction 
$f=\Omega_{\rm std}/\odm$ for bino-like neutralinos in the standard and non-standard cosmological models. The solid line indicates the CDMS II present limit~\cite{CDMSII}. The dashed lines show sensitivity limits for -from top to bottom on the right-  CDMS II,   ZEPLIN IV , XENON-1Ton,  and SuperCDMS phase C~\cite{Gaitskell:2004gd}.
}
\label{bin}
\end{center}
\end{figure}

Figure \ref{bin} displays the detection rate in standard and non-standard
cosmologies for bino-like neutralinos as a function of the neutralino mass. As before, the figure has been divided
into a rectangular grid and each occupied cell denotes the
existence of at least one model around it. For comparison, we also
show the current limit from the CDMS II experiment~\cite{CDMSII} as well as the expected
sensitivity of  CDMS II,   ZEPLIN IV , XENON-1Ton,  and SuperCDMS phase
C~\cite{Gaitskell:2004gd}. In the standard scenario, both the lower and the
upper limit on the bino mass are set by the relic density constraint. That is
why the range of neutralino masses extends to lower and higher values in
non-standard cosmologies. They yield  many more viable models, though most of them have rather
small detection rates. This fact is not entirely surprising. Small
annihilation rates, as those associated with bino-like neutralinos, are
generically correlated with small scattering rates. Regarding dark matter searches, the most remarkable difference observed in the
figure is the existence of new viable models with neutralino masses not
allowed in the standard cosmology and  detection rates  within the reach of future
experiments. Such models feature either $m_\chi\lesssim 40\gev$ or $m_\chi\gtrsim
600\gev$ and may be detected in ZEPLIN IV, XENON-1Ton, or SuperCDMS phase
C. 

\begin{figure}
\begin{center}
\includegraphics[scale=0.5]{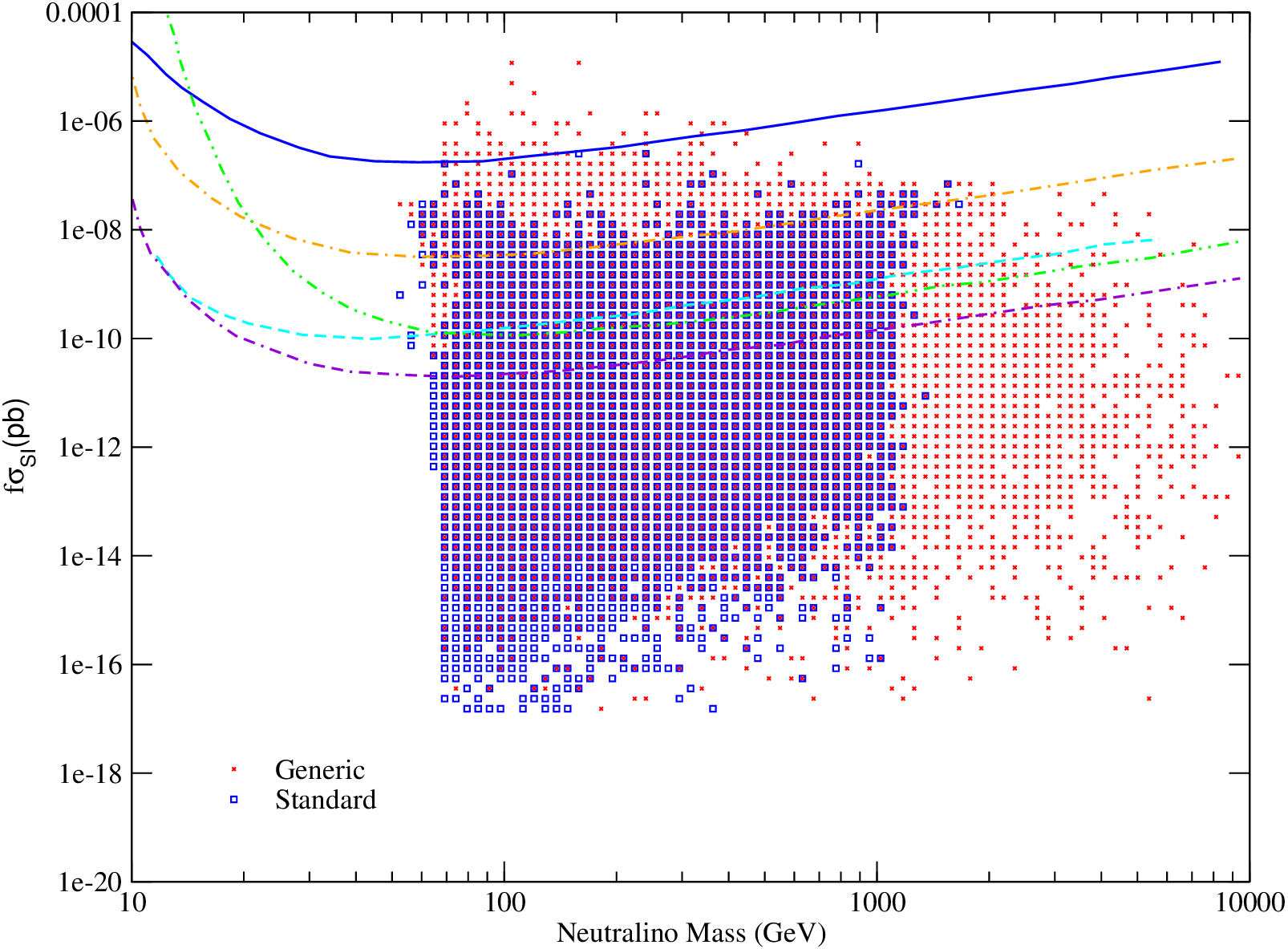}
\caption{
Spin-independent neutralino-proton cross section $\sigma_{SI}$ multiplied by the neutralino halo fraction 
$f=\Omega_{\rm std}/\odm$ for higgsino-like neutralinos in the standard and non-standard cosmological models. The solid line indicates the CDMS II present limit~\cite{CDMSII}. The dashed lines show sensitivity limits for -from top to bottom on the right-  CDMS II,   ZEPLIN IV , XENON-1Ton,  and SuperCDMS phase C~\cite{Gaitskell:2004gd}.}
\label{hig}
\end{center}
\end{figure}

The detection rate for higgsino-like neutralinos is shown in figure \ref{hig}
as a function of the neutralino mass in standard and non-standard
cosmologies. The lower limit on the higgsino mass
is now set by the experimental constraint on the chargino mass and is therefore
independent of the cosmological scenario. Two features clearly distinguish the
standard and the non-standard cosmologies. One of them is the existence of
viable models with heavy neutralinos, $m_\chi\gtrsim 1\tev$. A sizable
fraction of them has detection rates large enough to be observed in ZEPLINIV,
XENON-1Ton, or SuperCDMS phase C. The other feature is the significant enhancement in the
detection rate of neutralinos lighter than $\lesssim 1\tev$. In the standard
scenario, such neutralinos are usually underdense (see
figure \ref{omega-bwh-MSSM}) and have suppressed detection rates. From the
figure we see that non-standard cosmologies yield an enhancement of up to two
orders of magnitude for the neutralinos with the largest detection rates. Some
of them  are already ruled out by the present limit and many more will be within
the expected sensitivity of the CDMSII experiment.

\begin{figure}
\begin{center}
\includegraphics[scale=0.5]{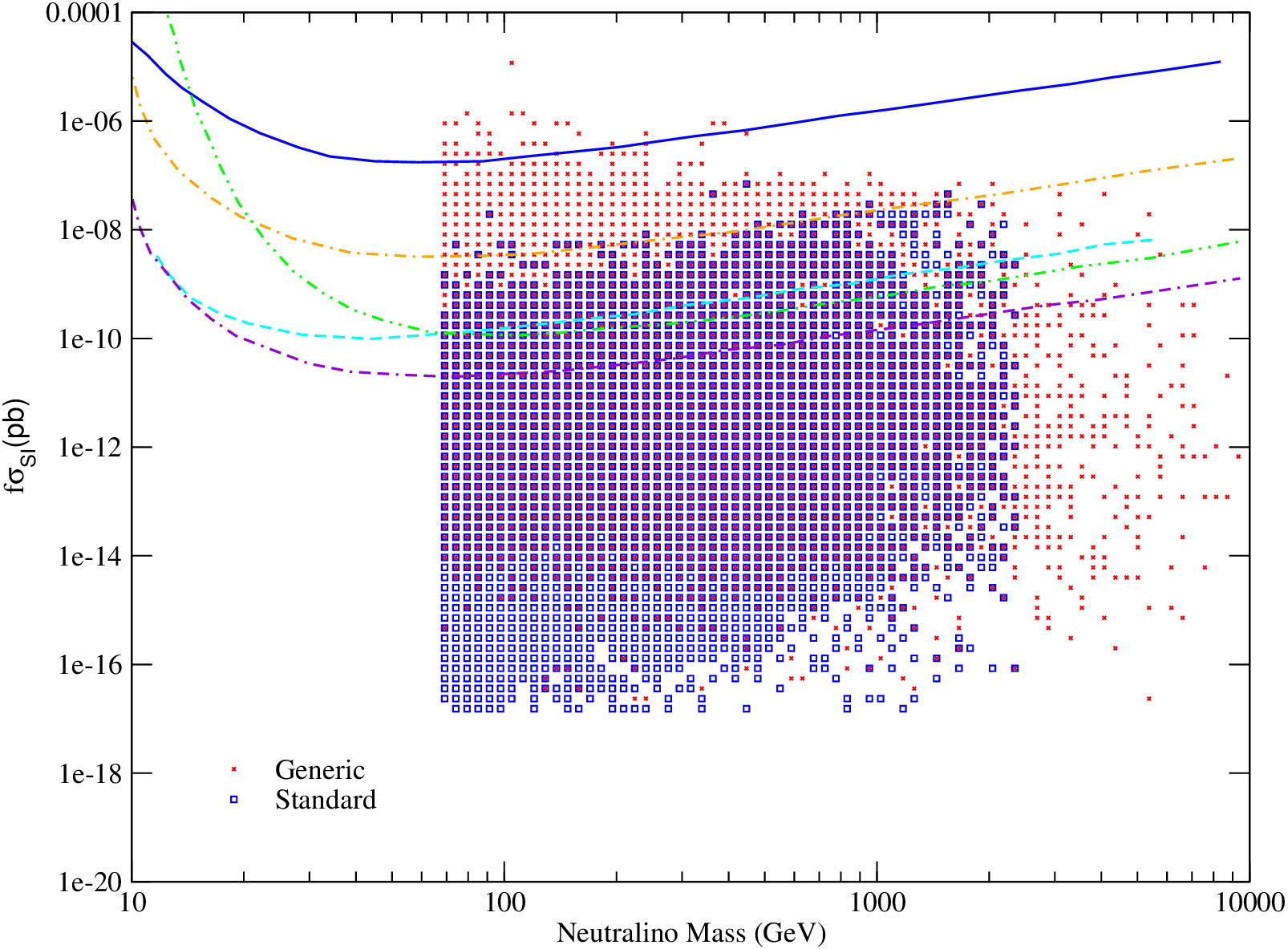}
\caption{
Spin-independent neutralino-proton cross section $\sigma_{SI}$ multiplied by the neutralino halo fraction 
$f=\Omega_{\rm std}/\odm$ for wino-like neutralinos in the standard and non-standard cosmological models. The solid line indicates the CDMS II present limit~\cite{CDMSII}. The dashed lines show sensitivity limits for -from top to bottom on the right-  CDMS II,   ZEPLIN IV , XENON-1Ton,  and SuperCDMS phase C~\cite{Gaitskell:2004gd}.
}
\label{win}
\end{center}
\end{figure}

A compelling signature of non-standard cosmologies would be the detection of a
wino-like neutralino by the CDMSII experiment, as revealed in figure
\ref{win}. Indeed, in the standard scenario,  winos with $m_\chi\lesssim 1\mbox{-}2\tev$ are
usually underdense and therefore their detection
rate is suppressed by the factor $f=\Omega_{\rm std}/\Omega_{\rm DM}$. In non-standard
cosmologies, such suppression is nonexistent and light winos have larger
detection rates. The enhancement in the
largest detection rates are typically larger than for higgsinos, amounting in
some cases to three orders of magnitude.  As for higgsinos, 
the lower bound on $m_\chi$ is not set by the dark
matter bound but rather by the experimental
constraint on the chargino mass, so no additional models are found at low
neutralino masses. For  $m_\chi\gtrsim 2\tev$ we do find
new viable models corresponding to overdense neutralinos in the standard cosmology. Most of
them, however, have small scattering rates, lying below the sensitivity of future detection experiments.   

\begin{figure}
\begin{center}
\includegraphics[scale=0.5]{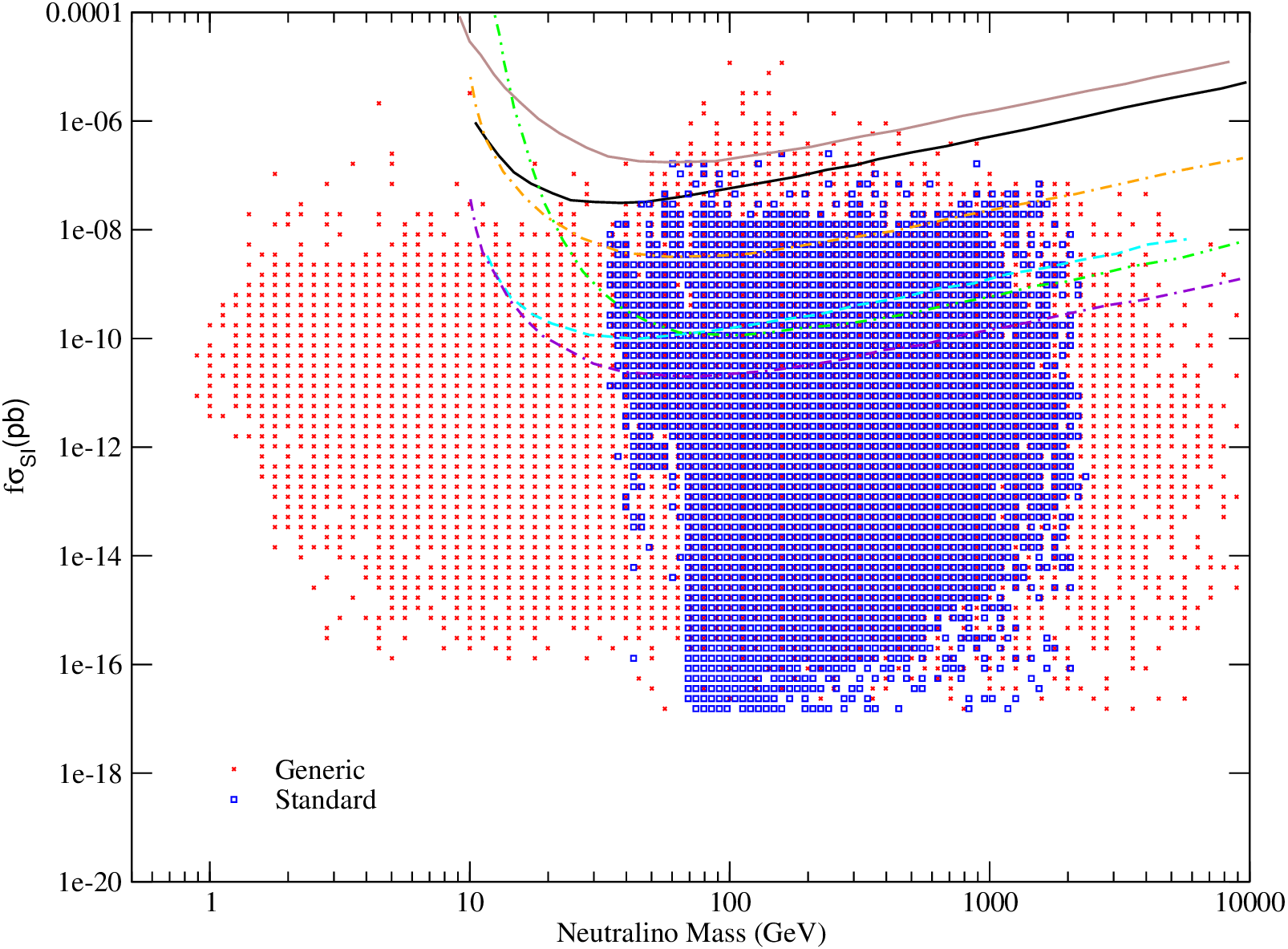}
\caption{
Spin-independent neutralino-proton cross section $\sigma_{SI}$ multiplied by the neutralino halo fraction  $f=\Omega_{\rm std}/\odm$  in the standard cosmological model and in the late decaying scalar field model. Here the lower limit of $M_1$ in Eq.~(2) has been lowered to 0.1 MeV
The solid upper line indicates the CDMS II present limit~\cite{CDMSII} and the lower solid line the XENON limit~\cite{XENON}. The dashed lines show sensitivity limits for -from top to bottom on the right-  CDMS II,   ZEPLIN IV , XENON-1Ton,  and SuperCDMS phase C~\cite{Gaitskell:2004gd}.}
\label{grac}
\end{center}
\end{figure}

{\textbf {Figure~\ref{grac} summarizes the potential increase in neutralino candidates in the models studied in references~\cite{gg} and \cite{ggsy}. For this figure the lower limit on $M_1$ in Eq.~(2) has been lowered to 0.1 GeV (which is compatible with all experimental limits (while no assumption is made on the relation between $M_1$ and $M_2$). In the late decaying scalar field scenario most neutrinos can be brought to have the dark matter density (provided the value of the two relevant parameters of the physics at the high scale can be suitably arranged). One exception is that of  very light neutralinos which would be very overdense in the standard cosmology. Requiring the reheating temperature to be above 
4 MeV~\cite{hannestad}, in order not to modify nucleosynthesis, from the equations of reference~\cite{gg}  it is immediate to see that  neutralinos of mass $m_\chi$ should have  a standard density
smaller than the dark matter density times $(m_\chi/ 120 {\rm MeV})^4$ for it to be possible to bring their density to be that of the dark matter in the late decaying scalar field scenario. This constraint is included in figure~\ref{grac} where it is clearly shown the increase in potential neutralino candidates in the particular non-standard cosmological model considered with respect to the standard cosmological model.}}

%
%

\section{Conclusion}
To summarize, in  this paper  we computed the  direct detection rate of MSSM
 neutralinos in generic cosmological scenarios where they constitute the dark
 matter of the Universe. When compared with the predictions of the standard
 cosmology, considerable differences were encountered. If the neutralino is
 bino-like, as in msugra models, additional light $m_\chi\lesssim 40\gev$ and
 heavy $m_\chi\gtrsim 600\gev$ neutralinos with non-negligible detection rates
 were found. They could be detected in a variety of dark matter experiments
 such as ZEPLINIV, XENON-1Ton, or SuperCDMS phase C. For higgsino-like
 neutralinos, we found enhancements of up to two orders of magnitude in the
 largest detection rates as well as new viable models with heavy
 $m_\chi\gtrsim 1\tev$ neutralinos. Both effects yielding detection rates  within the sensitivity of
 future experiments. Wino-like neutralinos provide the clearest signature of
 non-standard cosmologies. Their detection rates may be enhanced by up to
 three orders of magnitude and they could be detected in CDMSII. Thus, the
 prospects for the direct detection of neutralinos in non-standard cosmologies
 are significantly more promising than in the standard scenario.

\begin{acknowledgments}
We thank Oleg Kalashev for allowing us to use the \emph{graphreader} program.
G.G., A.S. and C.Y.  were supported in part by the US Department of Energy Grant
DE-FG03-91ER40662, Task C  and G.G. also by NASA grants NAG5-13399  and ATP03-0000-0057 at UCLA. P.G. was  supported  in part by  the NFS
grant PHY-0456825 at the University of Utah.
\end{acknowledgments}

\end{document}